\documentclass[a4paper,11pt]{amsart}
\usepackage{etoolbox}

\makeatletter
\patchcmd{\@setauthors}{\MakeUppercase}{}{}{}
\makeatother

\usepackage{multicol}
\usepackage{multirow}
\usepackage{amsmath,latexsym,amsbsy,amssymb,fancyhdr}
\usepackage{array}
\usepackage{enumerate}
\usepackage{amsthm}
\usepackage{booktabs} 
\usepackage{multirow}
\usepackage{tabularx}
 \usepackage{epsfig} 

\usepackage{psfrag,graphicx}
\usepackage{epstopdf}
\usepackage[latin1]{inputenc}
\usepackage[colorlinks=true]{hyperref}
\hypersetup{urlcolor=blue, citecolor=blue, linkcolor=blue}\usepackage{multicol}
\usepackage{multirow}
\usepackage{amsmath,latexsym,amsbsy,amssymb,fancyhdr}
\usepackage{array}
\usepackage{enumerate}
\usepackage{amsthm}
\usepackage{booktabs} 
\usepackage{multirow}
\usepackage{tabularx}
 \usepackage{epsfig} 

\usepackage{psfrag,graphicx}
\usepackage{epstopdf}
\usepackage[latin1]{inputenc}
\usepackage[colorlinks=true]{hyperref}
\hypersetup{urlcolor=blue, citecolor=blue, linkcolor=blue}

\usepackage{color}
\usepackage{cite}
\usepackage{commath}
\usepackage{mathtools}
\usepackage{systeme}
\usepackage[titletoc,toc,title]{appendix}

\usepackage{subfigure}

\newcommand{\at}[2][]{#1|_{#2}}

\numberwithin{equation}{section}

\makeatletter
 
\let\insertdate\@date
\makeatother

\usepackage{xcolor}
\definecolor{palegreen}{rgb}{0.2,0.6,0.2}

\author{Binh Duc Truong, Cuong Phu Le and Einar Halvorsen.}
\begin{document}
\title{On the lateral instability analysis of MEMS comb-drive electrostatic transducers}
\maketitle


\keywords{\textbf{\textit{Abstract --}} \small This paper investigates the lateral pull-in effect of an in-plane overlap-varying transducer. The instability is induced by the translational and rotational displacements. Based on the principle of virtual work, the equilibrium conditions of force and moment in lateral directions are derived. The analytical solutions of the critical voltage, at which the pull-in phenomenon occurs, are developed when considering only the translational stiffness or only the rotational stiffness of the mechanical spring. The critical voltage in general case is numerically determined by using nonlinear optimization techniques, taking into account the combined effect of translation and rotation. The effects of possible translational offsets and angular deviations to the critical voltage are modeled and numerically analyzed. The investigation is then the first time expanded to anti-phase operation mode and Bennet's doubler configuration of the two transducers.}

\section{Introduction}

The comb-drive electrostatic transduction is one of the most popular mechanism used in MEMS due to its many inherent advantages such as high efficiency and low power consumption. Various comb-drive electrostatic transducers have been early developed and utilized in wide variety of application, including micro energy harvesting \cite{Meninger2001, Roundy2002}, micro resonators \cite{Tang1990, Adams1998} and micro actuators \cite{Hirano1992, Jaecklin1992}. During operation, a voltage is applied to the device, generating an electrostatic force between fixed and movable electrodes, both in stroke direction and its perpendicular direction. 
At critical condition when the electrostatic force exceeds the mechanical restoring force, a small disturbance could lead to collapsing of the movable fingers to the fixed ones. This restriction is more critical when the MEMS transducer is electrically configured as Bennet's doubler or voltage multiplier \cite{Truong2017a, Truong2017b}.
Design of comb-drive devices therefore requires a comprehensive analysis of pull-in effect since the travel range and device performance are severely limited by the inherent instability.

The pioneering investigation of pull-in phenomenon were presented in the late 1960's by Nathanson \textit{et al.} \cite{Nathanson1967}, in which the electrostatic deflection of a parallel-plate actuator is modeled by use of mass-spring system. The maximum possible deflection is derived as one-third of the initial gap. Since then, the nature of pull-in instability has attracted more and more attention. Other than that, instead of focusing on analysis, several researchers turned their interest towards designing of mechanical spring structures to enlarge the maximum displacement or devising an external control scheme to ensure the device stabilization.

Legtenberg \textit{et al.} presented an expression for the translational instability voltage and deflection \cite{Legtenberg1996}. The theoretical stiffness of various spring structures such as clamped-clamped, crab-led and folded-beam designs were determined. A similar issue with a tiled folded-beam suspension was investigated by Zhou \textit{et al.} \cite{Zhou2003}. Both theoretical and experimental results show an enhancement of the stable travel range. In these works, the rotational displacement has not been concerned yet.

Pull-in effect due to translational and rotational misalignments are individually analyzed by Avdeev \textit{et al.} utilizing three approaches: analytical, uncoupled 2D/3D finite element (FE) models and coupled FE model \cite{Avdeev2003}. A good agreement between analytical solutions and coupled FE simulation results show that fringing fields have little effect on the translational pull-in voltage for the comb-drive geometry. The critical voltage (i.e., beyond which the lateral instability occurs) considering both the translational stiffness and the rotational stiffness has not been explored yet.

Huang \textit{et al.} presented a development of this analysis, taking into account the case when effects of the translational and rotational deflections are comparable \cite{Huang2004}. Simplified analytical solutions of the pull-in voltage are obtained based on a two-dimensional model of a single movable comb finger. An simple example with two-port actuator was analyzed, in which the mechanical stiffnesses were calculated using ANSYS and the critical voltage were specifically determined. However, the cross stiffness between the translation and the rotation is neglected.

With the same manner, in this work, we further develop comprehensive theoretical model to investigate the lateral side instability phenomena for both two-port and three-port transducers. Analytic and numerical results can be adapted to any mechanical spring structure. An analysis that takes into account the effect of a translational or rotational offset due to potential process errors is presented. In general case when the cross stiffness between the two degrees of lateral freedom is included, the critical voltage for different transducer configurations are numerically studied. Nonlinear optimization techniques with unequal constraints are used due to complexity of the problem, especially when the two transducers are electrically configured as Bennet's doubler.
A complete design is given as an example without compromising the generality of our study.

\section{Analytical model of a single transducer with translational and rotational misalignments}

\subsection{Device modeling}

\begin{figure}[!tbp]
	\centering
	\includegraphics[width=0.35\textwidth]{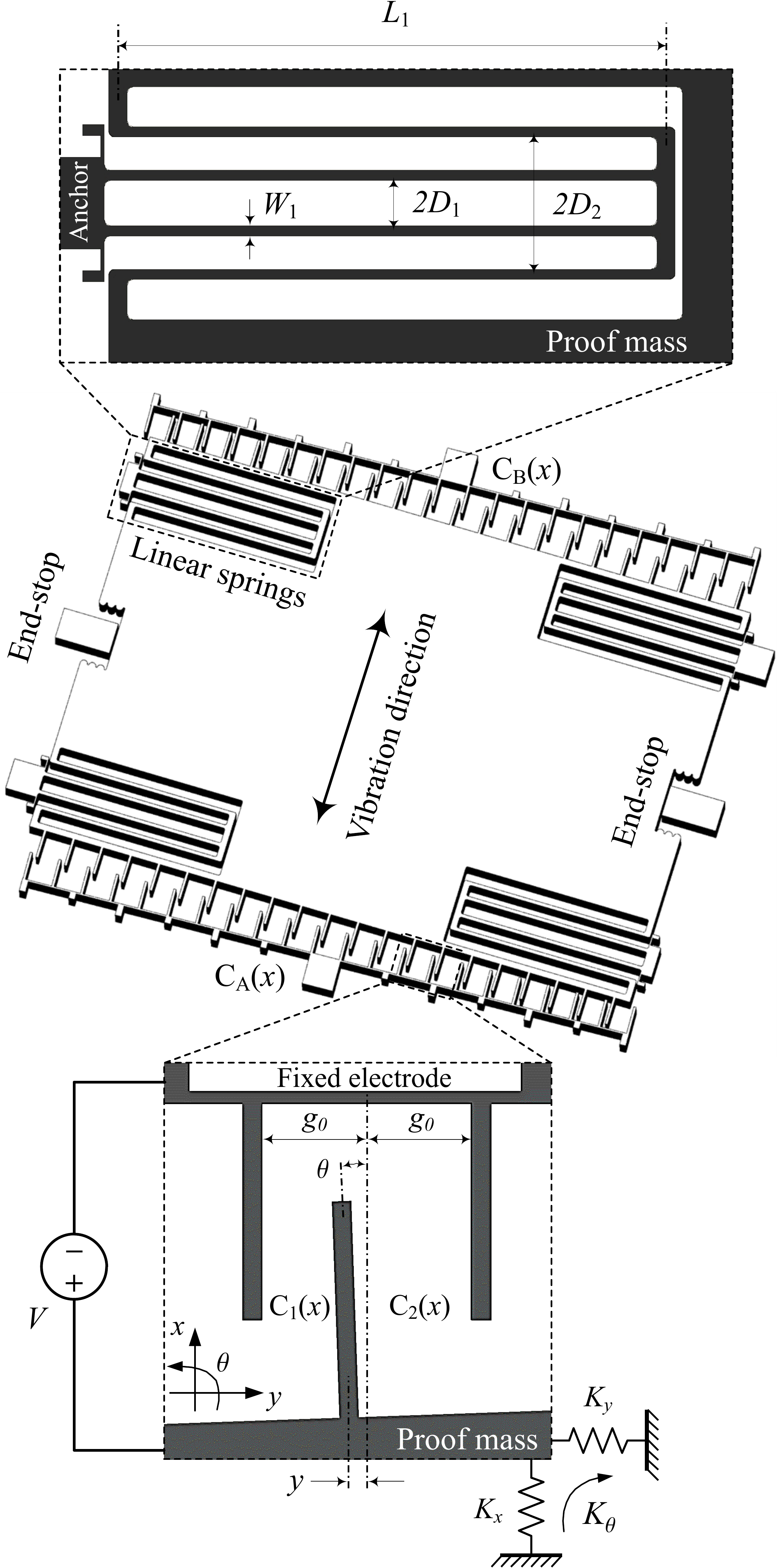}
	\caption{Key features of MEMS electrostatic transducers.}
	\label{Fig:Device}
\end{figure}
Figure \ref{Fig:Device} shows key features of the overlap-varying electrostatic transducers and addresses potential issues of the general lateral instability. The three-degree-of-freedom ($x,\, y,\, \theta$) device includes two ordinary comb-drive structures with proof mass suspended by four linear springs. The rigid end-stops are used to confine the maximum displacement. 
In an ideal case, the movable fingers are in the center of the gap along the $x$ axis, i.e. the stroke direction, and are in parallel with the fixed ones. 
The comb-drive fingers are assumed to be rigid due to the fact that their stiffness is typically designed to be much higher than the spring stiffnesses.

\begin{figure}[!tbp]
	\centering
	\includegraphics[width=0.12\textwidth]{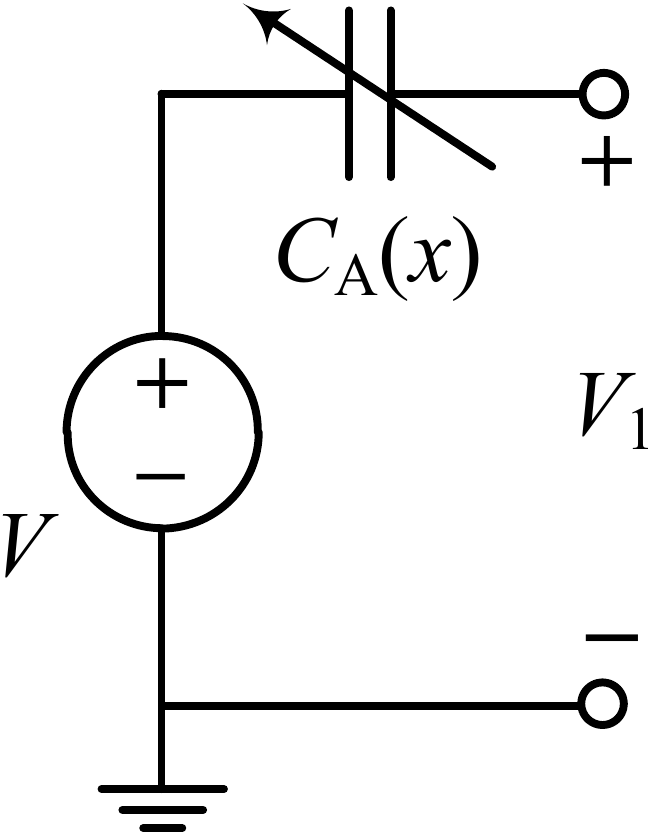}
	\caption{Circuit diagram for single variable capacitance device.}
	\label{Fig:Single_Trans}
\end{figure}
We are now considering the simplest case when a single electrostatic transducer is used as an actuator. Several prototypes were fabricated and evaluated, for instance, see among others \cite{Zhou2003, Grade2003, Chen2004, Olfatnia2013}.
An example of circuit diagram for this device type is drawn in Figure \ref{Fig:Single_Trans}.
As both the translational and rotational displacement are taken into account, i.e. $y$ and $\theta$ in close-up view of Figure \ref{Fig:Device}, capacitances of the transducer can be expressed
\begin{align}
\small
C_\mathrm{A} \big(x, y,  \theta \big) = C_1  \big(x, y,  \theta \big) + C_2  \big(x, y,  \theta \big) + C_\mathrm{p} \label{C_A_original}
\end{align}
where
\begin{align}
\small
C_1  \big(x, y,  \theta \big) &= N  \epsilon  \epsilon _0 t  \int_0^{x_0 + x}  \frac{1}{g_0 + y + (L-l) \sin \theta} \mathrm{d}l , \\
C_2  \big(x, y,  \theta \big) &= N  \epsilon  \epsilon _0 t  \int_0^{x_0 + x}  \frac{1}{g_0 - y - (L-l) \sin \theta} \mathrm{d}l  ,
\end{align}
$C_\mathrm{p}$ - the parasitic capacitance, $N$ - a number of the movable fingers, $\epsilon_0$ - the permittivity of free space, $\epsilon$ - the relative permittivity of the dielectric material, $t$ - the device thickness, $x_0$ - the initial overlap, $x$ - the proof mass displacement, $g_0$ - the initial gap between fingers, $L$ and $\mathrm{d}l$ - the length and a differential segment of the movable finger respectively. The fringing effect are ignored and the capacitance creating by the finger tips is negligible since the finger thickness is usually much smaller than its length.
These equation yields to
\begin{align}
\small
C_1  \big(x, y,  \theta \big) &= C_0  \frac{g_0}{2 x_0 \sin \theta} \ln \frac{g_0 + y + L \sin \theta}{g_0 + y +  \big(L -  \big(x_0 +x\big) \big)  \sin \theta}  \label{C1_gen} , \\
C_2  \big(x, y,  \theta \big) &= C_0  \frac{g_0}{2 x_0 \sin \theta} \ln \frac{g_0 - y - \big(L -  \big(x_0 +x\big) \big) \sin \theta}{g_0 - y - L \sin \theta} \label{C2_gen}
\end{align}
where $C_0 =  \frac{2N  \epsilon  \epsilon_0 t x_0}{g_0} $ is the nominal capacitance. Since the maximum displacement $X_\mathrm{max}$ is chosen to be smaller or equal the initial overlap, we get $x_0 \pm x \geq 0$, $ \forall x  \in [-X_\mathrm{max},\, X_\mathrm{max}] $.

\begin{figure}[!tbp]
	\centering
	\includegraphics[width=0.45\textwidth]{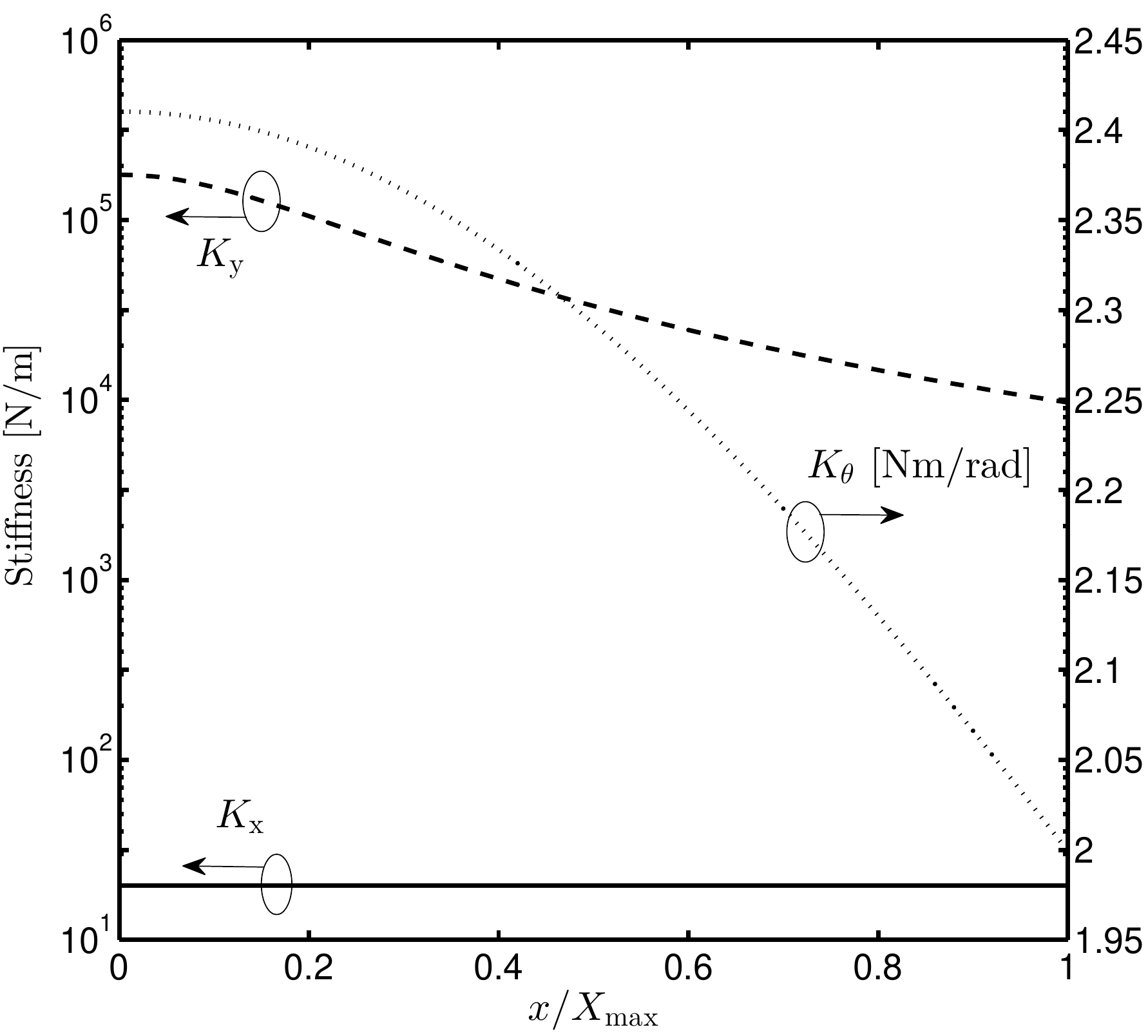}
	\caption{The displacement-dependent spring stiffnesses of the folded-beam flexure design.}
	\label{Fig:Stiffnesses}
\end{figure}
In this paper, we chose to investigate the folded-beam flexure as shown in Figure \ref{Fig:Device} which is one of the most commonly used suspensions in comb-drive transducers/actuators. Adapting from a work presented by Olfatnia \textit{et al.} \cite{Olfatnia2013} which included theoretical analysis and its experimental verification, the stiffness of a single spring are given
\begin{align}
\small
k_x &= \frac{Et W_1^3}{L_1^3} , \\
k_y &= \frac{Et W_1^3}{L_1} \frac{140}{140 W_1^2 + 51 x^2} , \\
k_\theta &= \frac{Et W_1}{L_1} \frac{350 W_1^2}{700 W_1^2 + 3 x^2} \frac{4 D_1^2 D_2^2}{D_1^2 + D_2^2}
\end{align}
where $E$ is Young's modulus. The spring length $L_1$, the spring width $W_1$ and the beam distances $D_1$ and $D_2$ are defined as Figure \ref{Fig:Device}. 

The total mechanical spring stiffnesses are $K_x = 4 k_x, K_y = 4 k_y$ and $K_\theta = 4 k_\theta$. It is important to observe that the translational and rotational stiffnesses $k_y$ and $k_\theta$ respectively decrease with the increase of the displacement $x$. In addition, $k_\theta$ can be made large with respect to the dimensions $D_1$ and $D_2$. Figure \ref{Fig:Stiffnesses} shows an analytical results of $K_y$ and $K_\theta$ in comparison with $K_x$. The drop in $K_y$ and $K_\theta$ from its nominal value (at $x=0$) with increasing $x$ is far more gradual. The detail parameters are summarized in Table \ref{Tab:Parameters}.

\begin{table}
	\small
	\centering
	\caption{Device structure parameters}%
	\begin{tabular}{l l} 
		\hline\hline
		\textbf{Parameters} & \textbf{Value} \\
		\hline
		Nominal capacitance, $C_0$ & 12.27 pF\\ 
		Device thickness, $t$ & 25 $\mu$m\\ 
		Finger length, $L$ & 222 $\mu$m\\
		Initial gap, $g_\mathrm{0}$ & 2 $\mu$m\\
		Nominal overlap, $x_\mathrm{0}$ & 110 $\mu$m \\
		Spring length, $L_\mathrm{1}$ & 1500 $\mu$m\\
		Spring width, $W_\mathrm{1}$ & 16 $\mu$m\\
		Beam distance, $D_1$ ($D_2$)  & 200 (90) $\mu$m \\
		Maximum displacement, $X_\mathrm{max}$ & 110 $\mu$m \\
		Young's modulus, $E$ & 169 GPa \\
		\hline\hline
	\end{tabular}
	\label{Tab:Parameters} 
\end{table}

\subsection{Potential energy}
For simplicity reasons, we only analyze the case $\abs x \leq X_\mathrm{max}$, the elastic energy of the end-stops is therefore neglected.
The total potential energy of the system can be written
\begin{align}
\small
W = W_\mathrm{m} + W_\mathrm{e}
\end{align}
where $W_\mathrm{m}$ is the elastic energy of the springs, $W_\mathrm{e}$ is the electrostatic energy of the transducers and $V$ is voltage applied to the electrodes. 
With the proof mass displaced by $x$ from the equilibrium position, their expressions are
\begin{align}
\small
W_\mathrm{m} &= \frac{1}{2} K_x x^2 + \frac{1}{2} K_y y^2 + \frac{1}{2} K_\theta \theta^2 , \\
W_\mathrm{e} &= -\frac{1}{2} \big(C_1 + C_2 + C_\mathrm{p}\big) V^2 .
\end{align}

According to the principle of virtual work, the forces and moment associated with the three coordinates $x, y$ and $\theta$ can be calculated by
\begin{align}
\small
F_\mathrm{x} &=  -\frac{\partial W}{\partial x} = - K_x x + \frac{1}{2} V^2 \frac{\partial \big(C_1 + C_2\big)}{\partial x} , \\
F_\mathrm{y} &=  -\frac{\partial W}{\partial y} = - K_y y + \frac{1}{2} V^2 \frac{\partial \big(C_1 + C_2\big)}{\partial y} , \\
M_\mathrm{\theta} &=  -\frac{\partial W}{\partial \theta} = - K_\theta \theta + \frac{1}{2} V^2 \frac{\partial \big(C_1 + C_2\big)}{\partial \theta} .
\end{align}
These forces and moment above characterize the equilibrium condition between the electrostatic forces and the restoring forces produced by the mechanical springs. The transducers are in the state of a static electromechanical equilibrium once all of them are equal to zero.
For a constant voltage, the transducers always seek out the orientation with the lowest potential energy.
If the equilibrium state corresponds to a local minimum of the potential $W$ then it is locally stable.
A local maximum or a saddle in potential energy corresponds to an equilibrium that is unstable.

\begin{figure}[!tbp]
	\begin{center}
		\subfigure[$W = W(y)$]{%
			\includegraphics[width=0.45\textwidth]{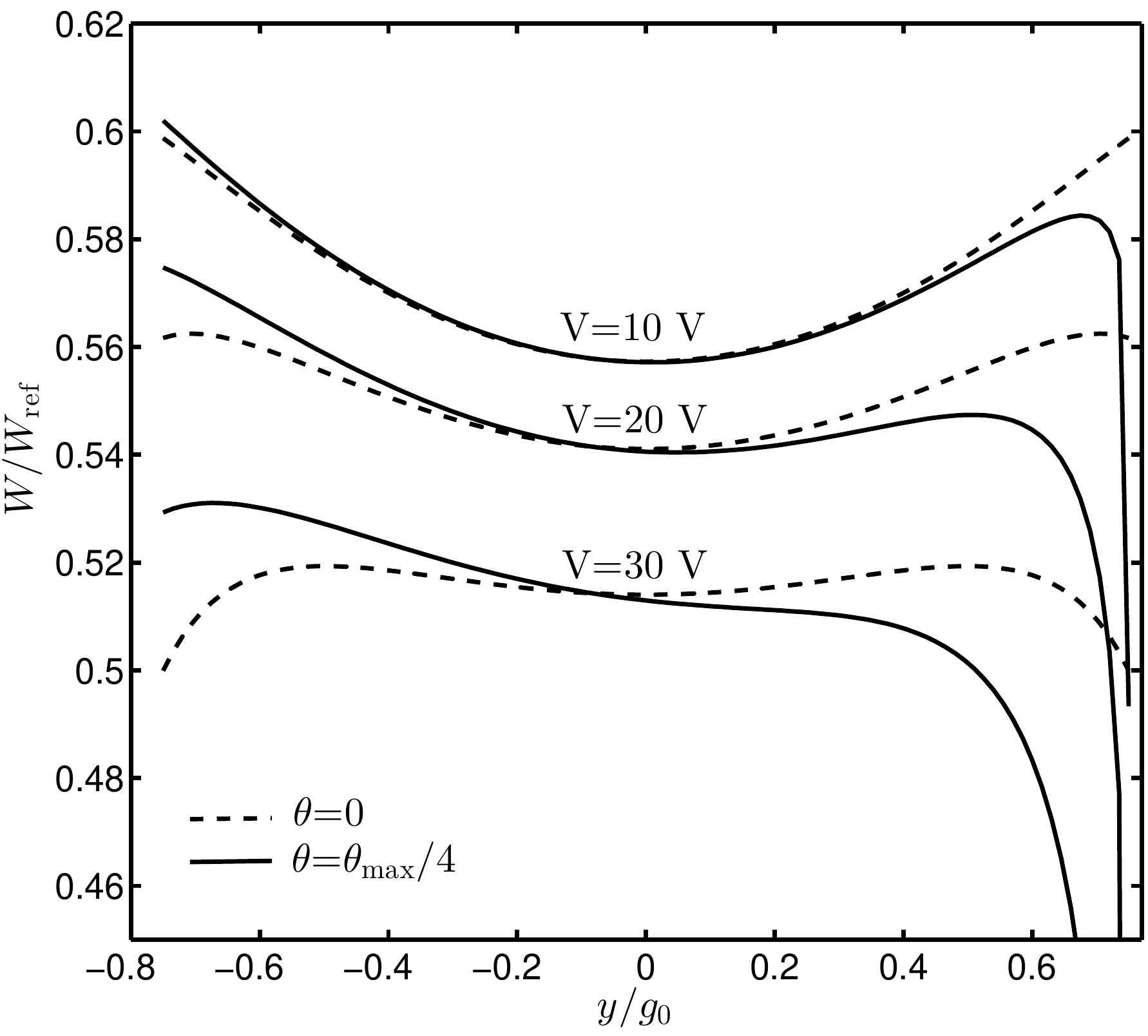}
		}%
		\space
		\subfigure[$W = W(\theta)$]{%
			\includegraphics[width=0.45\textwidth]{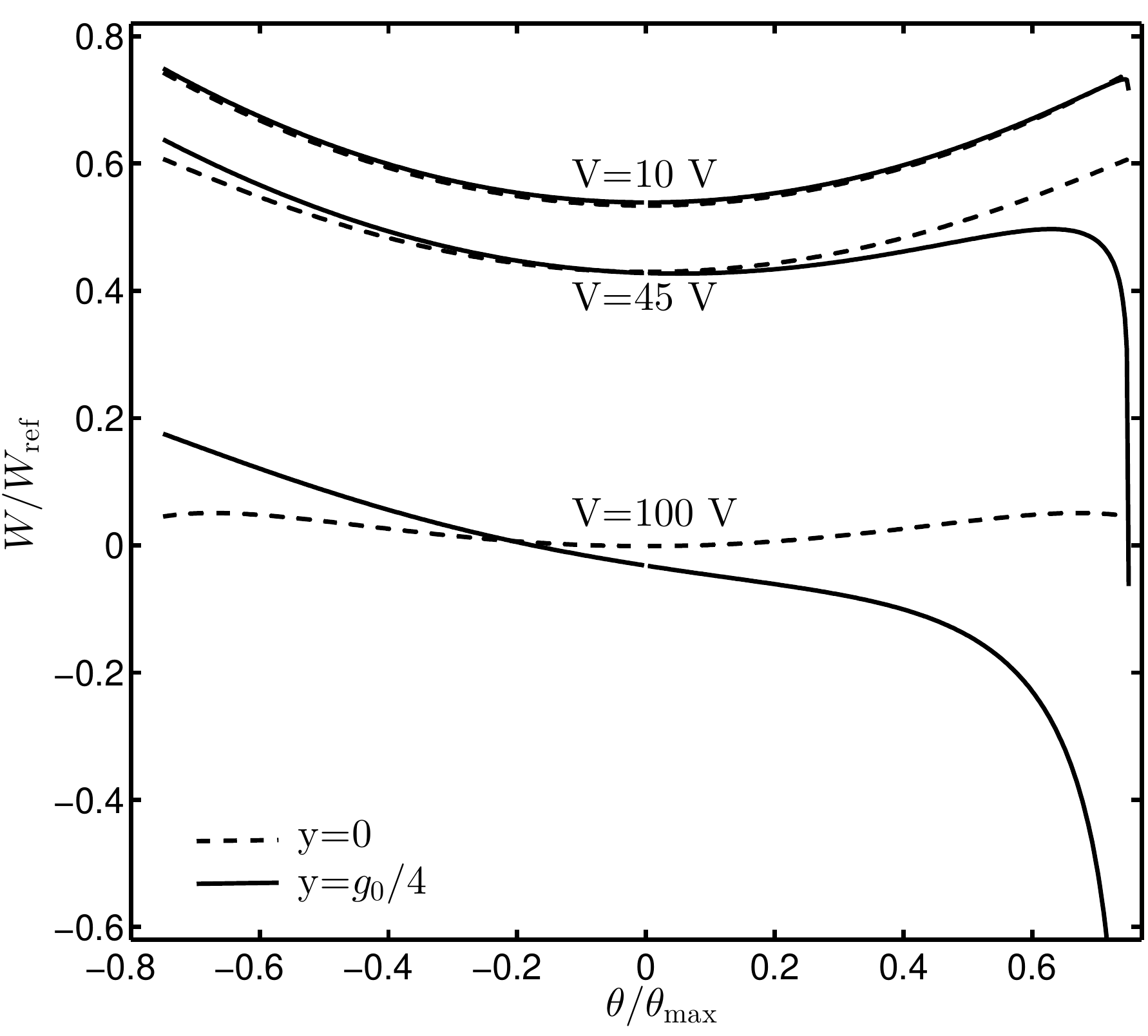}
		}\\ 
		\subfigure[$W = W(y,\,\theta)$]{%
			\includegraphics[width=0.45\textwidth]{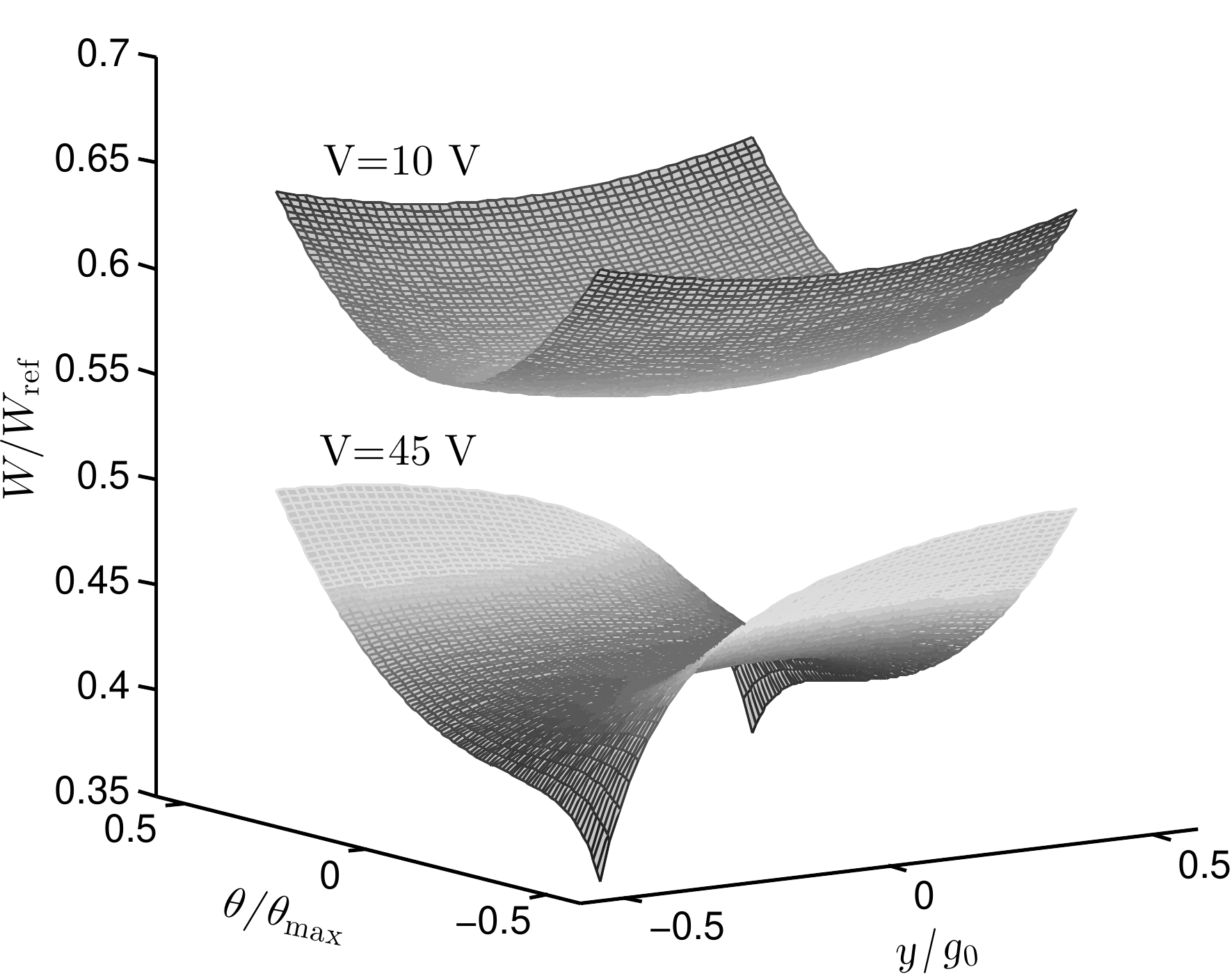}
		}%
	\end{center}
	\caption{Total potential energy of the transducers as a function of (a) the translational displacement $y$, (b) the rotational displacement $\theta$ and (c) both of $y$ and $\theta$, where $\theta_\mathrm{max} = \tan ^{-1} \frac{g_0}{L}$, $x = X_\mathrm{max}$ and $W_\mathrm{ref} = \frac{1}{2} K_x x^2$.}
	\label{Fig:Potential_En}
\end{figure}
Figure \ref{Fig:Potential_En} shows the total potential energy of the transducers at $x=X_\mathrm{max}$ and effect of the lateral translational and rotational displacement on the instability. For instance, considering $W$ as a function of $y$ only, i.e. Figure \ref{Fig:Potential_En} (a), in the case of $V=30$ V and $\theta=0$, the equilibrium state $y=0$ is stable as it is a local minimum of $W$. If $\theta=\frac{\theta_\mathrm{max}}{4}$, the equilibrium becomes unstable with the same voltage $V$ and any small perturbation of $y$ leads to the pull-in effect. In Figure \ref{Fig:Potential_En} (b), the same phenomenon happens with $V=100$ V and $y=\frac{g_\mathrm{0}}{4}$ as $W$ is a function of $\theta$ alone. Figure \ref{Fig:Potential_En} (c) provides us a more general evaluation of $W$ when different values of the constant voltage are applied. There is one stable equilibrium with $V=10$ V while those that of $V=45$ V are unstable. The transducer can exhibit equilibria that are unstable along the $y-$ or $\theta-$direction only or along both the $y-$ and $\theta-$directions.
In the following sections, analytical solution of the critical voltage when considering the rotational instability or the translational instability is developed. Numerical solution taking into account the combined effect of rotation and translation is investigated.

\subsection{Rotational instability}

Assuming that the translational stiffness $K_y$ is extremely large and the translation $y$ is negligible. Since $\theta$ is small, $\sin \theta \approx \theta$, the total capacitance is
\begin{align}
\small
C_\mathrm{A} = \frac{1}{2} \frac{C_0 g_0}{x_0 \theta} \ln \frac{\big(g_0 + L \theta \big) \big(g_0 - \big( L - \big(x_0 + x\big) \big) \theta \big)}{\big(g_0 - L \theta \big) \big( g_0 + \big(L -  \big(x_0 +x\big) \big)  \theta \big)} + C_\mathrm{p}\label{C_A} .
\end{align}
The capacitance changes with the stroke direction displacement $x$ and rotational angle $\theta$. When the moving fingers are parallel to the fixed ones (i.e., $\theta = 0$), the capacitance simplifies to the well-know parallel-plate calculation as expected $\displaystyle{\lim_{\theta \to 0} C_\mathrm{A} = C_0 \big(1 + \frac{x}{x_0}\big) + C\mathrm{p}}$.
However, as will be shown later, this does not indicate that the rotation effect can be neglected.

At equilibrium, the electrostatic moment is balanced by that of the mechanical spring, which implies
\begin{align}
\small
\frac{\partial^2 W}{\partial \theta^2} = -\frac{\partial M_\mathrm{\theta}}{\partial \theta} \at[\bigg]{\theta  \rightarrow 0} = K_\theta -  \frac{1}{3} V^2 \frac{C_0}{x_0}  \frac{ \big(x_0 + x\big)  \big(3L^2 - 3L \big(x_0 + x\big)  + \big(x_0 + x\big)^2\big)  }{g_0 ^2} = 0 .
\end{align}
The requirement for stability is that the potential energy is concave up, or equivalently $\frac{\partial^2 W}{\partial \theta^2} > 0$.
The maximum voltage across the transducer, so-called critical (or pull-in) voltage, to avoid lateral instability due to rotation is given as
\begin{align} \label{V_theta_cr_full}
\small
V_\mathrm{\theta-cr}= \sqrt{3  \frac{g_0 ^2 x_0}{C_0}  \frac{K_\mathrm{\theta}}{\big(x_0 + x\big)  \big(3L^2 - 3L \big(x_0 + x\big)  + \big(x_0 + x\big)^2\big)} } .
\end{align}
If the voltage is greater than $V_\mathrm{\theta-cr}$, the transducer cannot be in equilibrium, and within certain time, the moving electrode will snap against the fixed one.

In case of $x=X_\mathrm{max} \approx x_0 \approx \frac{L}{2}$, equation \eqref{V_theta_cr_full} yields
\begin{align} \label{V_theta_cr_Xmax}
\small
V_\mathrm{\theta-cr} \at[\Big]{x=X_\mathrm{max}}= \sqrt{\frac{3}{2} \frac{g_0 ^2}{C_0} \frac{K_\mathrm{\theta}}{L^2}} .
\end{align}
It is obvious that increase of the nominal gap $g_0$ enhances the lateral stability. However, on the other hand, some device functions (e.g., energy harvesters) may require large nominal capacitance $C_0 =  \frac{2N  \epsilon  \epsilon_0 t x_0}{g_0} $. This could perhaps lead to decrease of $g_0$ especially when the initial overlap $x_0$ is limited. Thus, design of a mechanical suspension with large $K_\theta$ would seem to be the more common point of view to increase the side stability.

As seen in equation \eqref{C_A_original}, the transducer capacitance is modeled by the ideal-capacitance plus the constant, parallel parasitic capacitance $C_\mathrm{p}$. Under voltage control, the derivatives of the capacitance are functions of displacements, i.e. do not contain $C_\mathrm{p}$ anymore. The pull-in voltage is therefore independent of $C_\mathrm{p}$. The relationship between charge and displacement in equilibrium that depends on $C_\mathrm{p}$ is out of the scope in this paper. In following sections, $C_\mathrm{p}$ will be eliminated.

\subsection{Translational instability}

As the rotational stiffness is large enough, the rotation can be neglected. Evaluating $\displaystyle{\lim_{\theta \to 0} (C_1 + C_2)}$ yields
\begin{align}
\small
C_\mathrm{A} = \frac{C_0 g_0^2 \big(x_0 + x\big)}{x_0} \frac{1}{g_0^2 - y^2} .
\end{align}
The static equilibrium condition is satisfied when
\begin{align}
\small
\frac{\partial^2 W}{\partial y^2} = \frac{\partial F_\mathrm{y}}{\partial y} \at[\bigg]{\substack{\theta  \rightarrow 0\\ y  \rightarrow 0}} = - K_y +  V^2 \frac{C_0 \big(x_0 + x\big)}{x_0 g_0^2}= 0 .
\end{align}
From which, the displacement-dependent critical voltage can be extracted
\begin{align} \label{V_y_cr_full}
\small
V_\mathrm{y-cr} = \sqrt{\frac{x_0 g_0 ^2}{C_0 \big(x_0 + x\big)} K_y} = \sqrt{\frac{g_0 ^3}{2N  \epsilon  \epsilon_0 t \big(x_0 + x\big)} K_y} .
\end{align}
Based on particular applications of the transducer, one should reasonably expect to make a trade-off between the nominal capacitance $C_0$ and the initial overlap $x_0$. For an example, in case that is to maximize the travel range while $V_\mathrm{y-cr}$ is kept the same, a design of the comb-drive device should have $x_0 = 0$ (or very small), however, yielding to $C_0 = 0$.

Similarly, at the maximum displacement, equation \eqref{V_y_cr_full} is simplified as
\begin{align} \label{V_y_cr_Xmax}
\small
V_\mathrm{y-cr}\at[\Big]{x=X_\mathrm{max}} = \sqrt{\frac{1}{2}\frac{g_0 ^2}{C_0 } K_y} .
\end{align}

\begin{figure}[!tbp]
	\centering
	\includegraphics[width=0.45\textwidth]{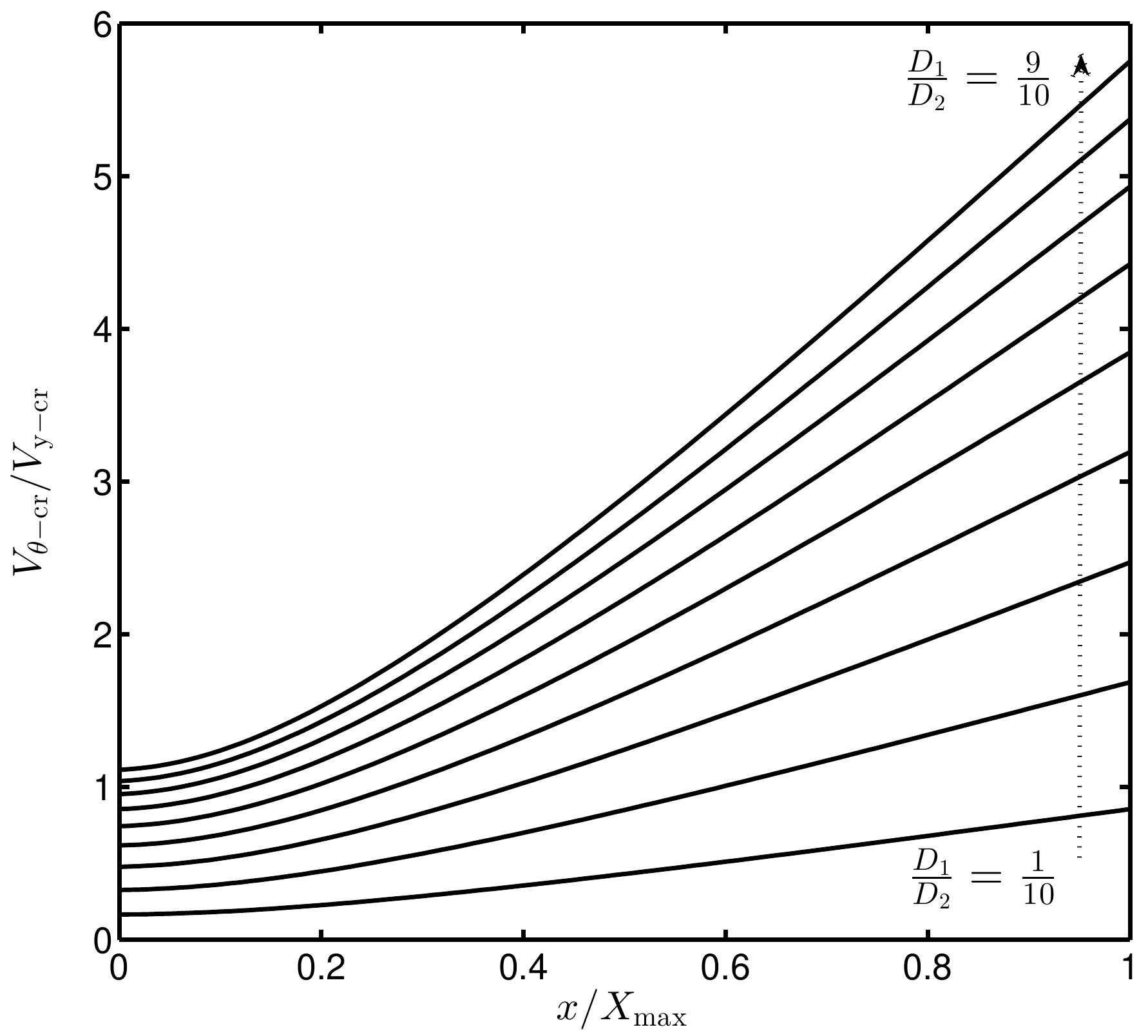}
	\caption{The ratio of rotational and translational critical voltages versus proof mass displacement with different values of $\frac{D_1}{D_2}$}
	\label{Fig:V_cr_ratio}
\end{figure}
Equations \eqref{V_theta_cr_Xmax} and \eqref{V_y_cr_Xmax} show that the ratio of these two critical voltages is proportional to root square of the corresponding stiffnesses
\begin{align} \label{V_cr_ratio}
\small
V_\mathrm{r-cr} = \frac{V_\mathrm{\theta-cr}}{V_\mathrm{y-cr}}  \propto \sqrt{\frac{K_\theta}{K_y}} \propto \sqrt{\frac{4 D_1^2 D_2^2}{D_1^2 + D_2^2}} .
\end{align}
The value of $V_\mathrm{r-cr}$ can be displacement-independently made large by appropriate choices of the dimensions $D_1$ and $D_2$.
Figure \ref{Fig:V_cr_ratio} depicts the variation of $V_\mathrm{r-cr}$ with respect to $x$, in which $V_\mathrm{r-cr}$ increases along with the increases of the ratio $\frac{D_1}{D_2}$.
Effect of the lateral rotation on the device instability is therefore markedly diminished if $\frac{D_1}{D_2}$ is large enough. For instance, $V_\mathrm{\theta-cr}$ is about 3.8 times higher than $V_\mathrm{y-cr}$ if $D_1=\frac{1}{2}D_2$. In this case, the lateral translation is more critical.

\subsection{Lateral instability due to combination of translation and rotation}

In general, when both the translational and rotational displacement is comparable, a stiffness matrix corresponding to the coordinates $y$ and $\theta$ contains a cross-interaction terms, i.e. $K_{y \theta}$ and $K_{\theta y}$. So far, however, all analyses of the lateral instability of the in-plane comb-drive MEMS transducers have been limited to neglect of the cross stiffness terms. In this paper, a further developed model taking into account the effect of $K_{y \theta}$ and $K_{\theta y}$ is explored. The moment and force equilibrium conditions now are
\begin{align} \label{Eq:Matrix}
\begin{bmatrix}
F \\M 
\end{bmatrix}
=
\begin{bmatrix}
\frac{\partial F_\mathrm{y}}{\partial y} \at[\Big]{y \rightarrow 0} & \frac{\partial F_\mathrm{y}}{\partial \theta} \at[\Big]{\theta  \rightarrow 0} \\ 
\frac{\partial M_\mathrm{\theta}}{\partial y} \at[\Big]{y  \rightarrow 0} & \frac{\partial M_\mathrm{\theta}}{\partial \theta} \at[\Big]{\theta  \rightarrow 0}
\end{bmatrix}
\begin{bmatrix}
y \\ \theta 
\end{bmatrix}
=
\begin{bmatrix}
0 \\ 0 
\end{bmatrix}
\end{align}
where the stiffness coefficients are given by
%
%
%
%
\begin{align}
\small
\begin{split}
\frac{\partial F_\mathrm{y}}{\partial y} \at[\Big]{y \rightarrow 0} = -K_y + \frac{1}{4} V^2 \frac{C_0 g_0}{x_0} \left[ \frac{4 g_0 L}{\big(g_0 + L \theta\big)^2 \big(g_0 - L \theta\big)^2} \right. \\
\left. + \frac{1}{\theta \big( g_0 + \big( L - (x_0 + x) \big) \theta \big)^2} - \frac{1}{\theta \big( g_0 - \big( L - (x_0 + x) \big) \theta \big)^2} \right] ,
\end{split}
\end{align}
\begin{align}
\small
\frac{\partial F_\mathrm{y}}{\partial \theta} \at[\Big]{\theta  \rightarrow 0} = \frac{1}{2} V^2 \frac{C_0 g_0^2}{x_0} \frac{\big(g_0^2 + 3y^2\big) \big(x_0 + x\big) \big( 2L - (x_0 + x)\big)}{\big(g_0 - y\big)^3 \big(g_0 + y\big)^3} ,
\end{align}
\begin{align}
\small
\frac{\partial M_\mathrm{\theta}}{\partial y} \at[\Big]{y  \rightarrow 0} = \frac{1}{2} V^2 \frac{C_0 g_0^2 (x_0 + x)}{x_0} \left[ \dfrac{\splitdfrac{ g_0^4 \big( 2L - (x_0 +x)\big) - 3L^2 \theta^4 \big( 2L - (x_0 +x)\big) \big( L - (x_0 +x)\big)^2 }{ + g_0^2 \theta^2 \big( 4L^3 - 6L^2 (x_0 + x) + 4L(x_0+x)^2 -(x_0 + x)^3 \big) }} {\big(g_0 - L \theta \big)^2 \big(g_0 + L \theta\big)^2} \right] ,
\end{align}
\begin{align}
\small
\frac{\partial M_\mathrm{\theta}}{\partial \theta} \at[\Big]{\theta  \rightarrow 0} = - K_\theta + \frac{1}{3} V^2 \frac{C_0 g_0^2}{x_0} \frac{\big(g_0^2 + 3y^2\big) \big(x_0 + x\big) \big( 3L^2 - 3L(x_0+x) + (x_0+x)^2 \big)}{\big(g_0 - y\big)^3 \big(g_0 + y\big)^3} .
\end{align}

Let $\overline{V}$ be a set of the parameter $V$ such that the equation \eqref{Eq:Matrix} has solutions $y \in D_\mathrm{y} \,\, \mathrm{and} \,\, \theta \in D_\mathrm{\theta}$, the critical voltage at specific proof mass position is expressed as
\begin{align}
\small
V_\mathrm{y, \theta - cr} = \displaystyle \max \left \{ V \in \overline{V} \right \}
\end{align}
where $D_\mathrm{y} : \left \{ \abs y < g_0 \right \}$ and $D_\mathrm{\theta} : \left \{ \abs \theta < \theta_\mathrm{max}=\tan^{-1} \frac{g_0}{L} \right \}$. 

\begin{figure}[!tbp]
	\centering
	\includegraphics[width=0.45\textwidth]{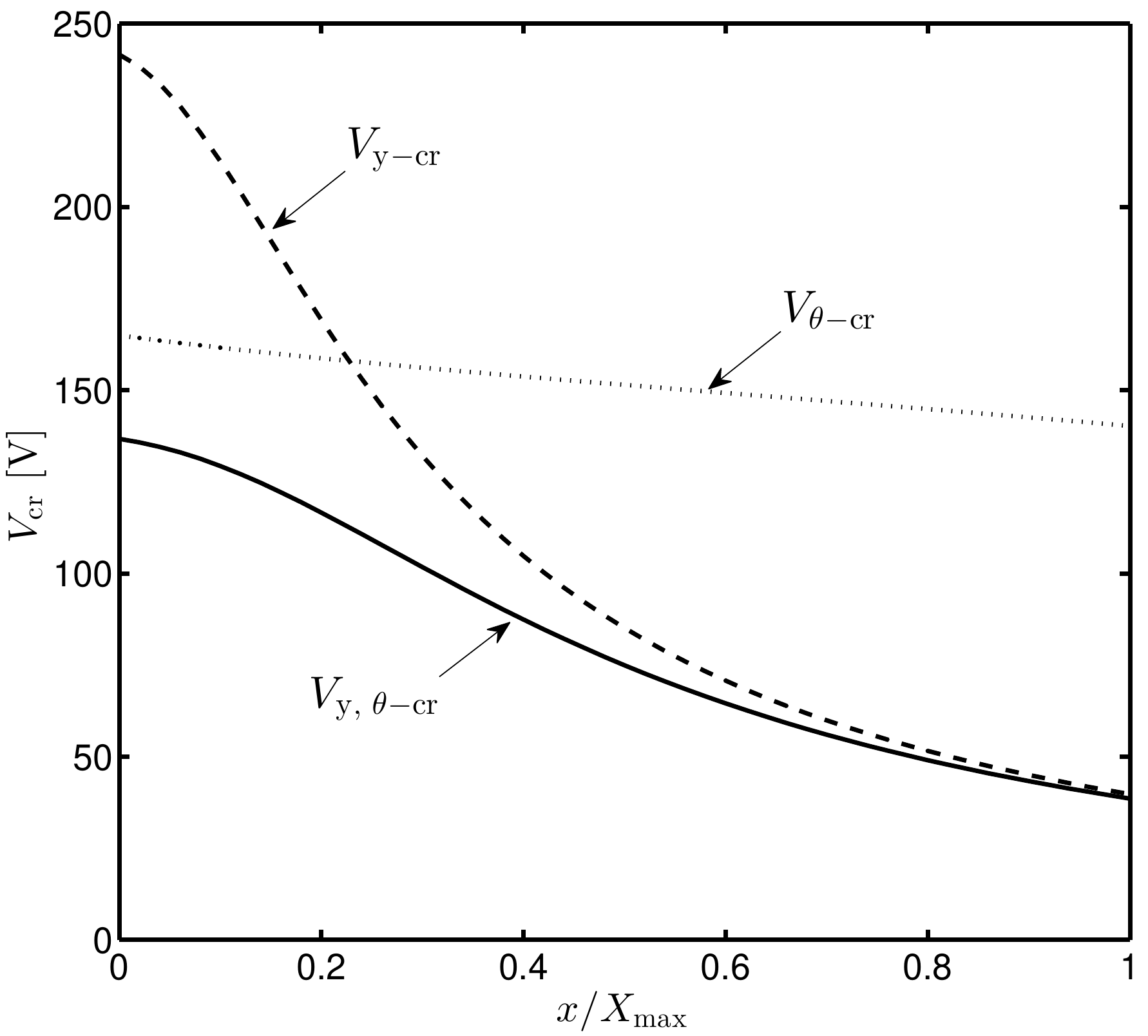}
	\caption{Comparison of the critical voltages in three analyzed cases: rotational instability $V_\mathrm{\theta-cr}$ or translational instability $V_\mathrm{y-cr}$, alone or in combination $V_\mathrm{y,\theta-cr}$}
	\label{Fig:V_cr}
\end{figure}
In order to solve such nonlinear optimization problem with the strict constrained conditions of $y$ and $\theta$, the nonlinear Interior Point (IP) or Sequential Quadratic Programming (SQP) methods are utilized \cite{Nocedal2006}. The numerical result of $V_\mathrm{y, \theta - cr}$ is compared to the analytical solutions of $V_\mathrm{\theta - cr}$ and $V_\mathrm{y - cr}$ obtained from equations \eqref{V_theta_cr_full} and \eqref{V_y_cr_full} respectively in Figure \ref{Fig:V_cr}. Obviously, the critical voltage considering both the translational and rotational displacements is always less than those considering one of them. For the folded beam suspension mechanism studied in this paper, $V_\mathrm{y - cr}$ dramatically drops while $V_\mathrm{\theta - cr}$ only slightly (and linearly) decreases when $x$ approaches its maximum $X_\mathrm{max}$. In this case, the effect of lateral translation should be more concerned since $V_\mathrm{y - cr}$ is much close to $V_\mathrm{y, \theta - cr}$ than $V_\mathrm{\theta - cr}$ at extreme position of the proof mass.

\subsection{Critical voltage with translational and rotational offsets}

Normally, for the overlap-varying electrostatic transducers, the movable fingers are placed in the between fixed ones. However, there is a possibility that exist translational and rotational offsets, i.e. $y_0$ and $\theta_0$ respectively, due to manufacturing tolerance or error in fabrication process. This can lead to further reduction of the critical voltage. The value of $V_\mathrm{y,\theta - cr}$ at $x=X_\mathrm{max}$ is investigated since it is the maximum voltage that can be applied between the two electrodes while still ensuring the transducer stability.

When $x=X_\mathrm{max} \approx x_0 \approx \frac{L}{2}$, the moment and force equilibrium conditions in equation \eqref{Eq:Matrix} becomes
\begin{align} \label{Eq:Matrix_Offsets}
\small
\begin{bmatrix}
-K_y + \frac{1}{2} V^2 A & \frac{1}{2} V^2 B \\ 
\frac{1}{2} V^2 C & -K_\theta + \frac{1}{2} V^2 D
\end{bmatrix}
\begin{bmatrix}
y_0 + \Delta y \\ \theta_0 + \Delta \theta 
\end{bmatrix}
=
\begin{bmatrix}
0 \\ 0 
\end{bmatrix}
\end{align}
where
\begin{align}
\small
A &= \frac{2 C_0 g_0^2 L}{x_0 \big(g_0 + L (\theta_0 + \Delta\theta)\big)^2 \big(g_0 - L (\theta_0 + \Delta\theta) \big)^2} , \\
B &= \frac{C_0 g_0^2 L^2 \big(g_0^2 + 3(y_0 + \Delta y)^2\big)}{\big(g_0 - (y_0 + \Delta y)\big)^3 \big(g_0 + (y_0 + \Delta y)\big)^3} , \\
C &= \frac{C_0 L^2 \big(g_0^2 + L^2 (\theta_0 + \Delta \theta)^2\big)}{x_0 \big(g_0 - L (\theta_0 + \Delta\theta)\big)^2 \big(g_0 + L (\theta_0 + \Delta\theta)\big)^2} , \\
D &= \frac{2}{3} \frac{C_0 g_0^2 \big(g_0^2 + 3(y_0 + \Delta y)^2\big) L^3}{x_0 \big(g_0 - (y_0 + \Delta y)\big)^3 \big(g_0 + (y_0 + \Delta y)\big)^3} .
\end{align}
%
%
%
%
The critical voltage is the intersection of two surfaces determined by
\begin{equation} \label{V_cr_Xmax}
\small
\left\{
\begin{aligned}
V &= \sqrt{\frac{2 K_y (y_0 + \Delta y)}{A (y_0 + \Delta y) + B (\theta_0 + \Delta \theta)}} \\
V &= \sqrt{\frac{2 K_\theta (\theta_0 + \Delta\theta)}{C (y_0 + \Delta y) + D  (\theta_0 + \Delta\theta) }}
\end{aligned}
\right.
\end{equation}

\begin{figure}[!tbp]
	\centering
	\includegraphics[width=0.45\textwidth]{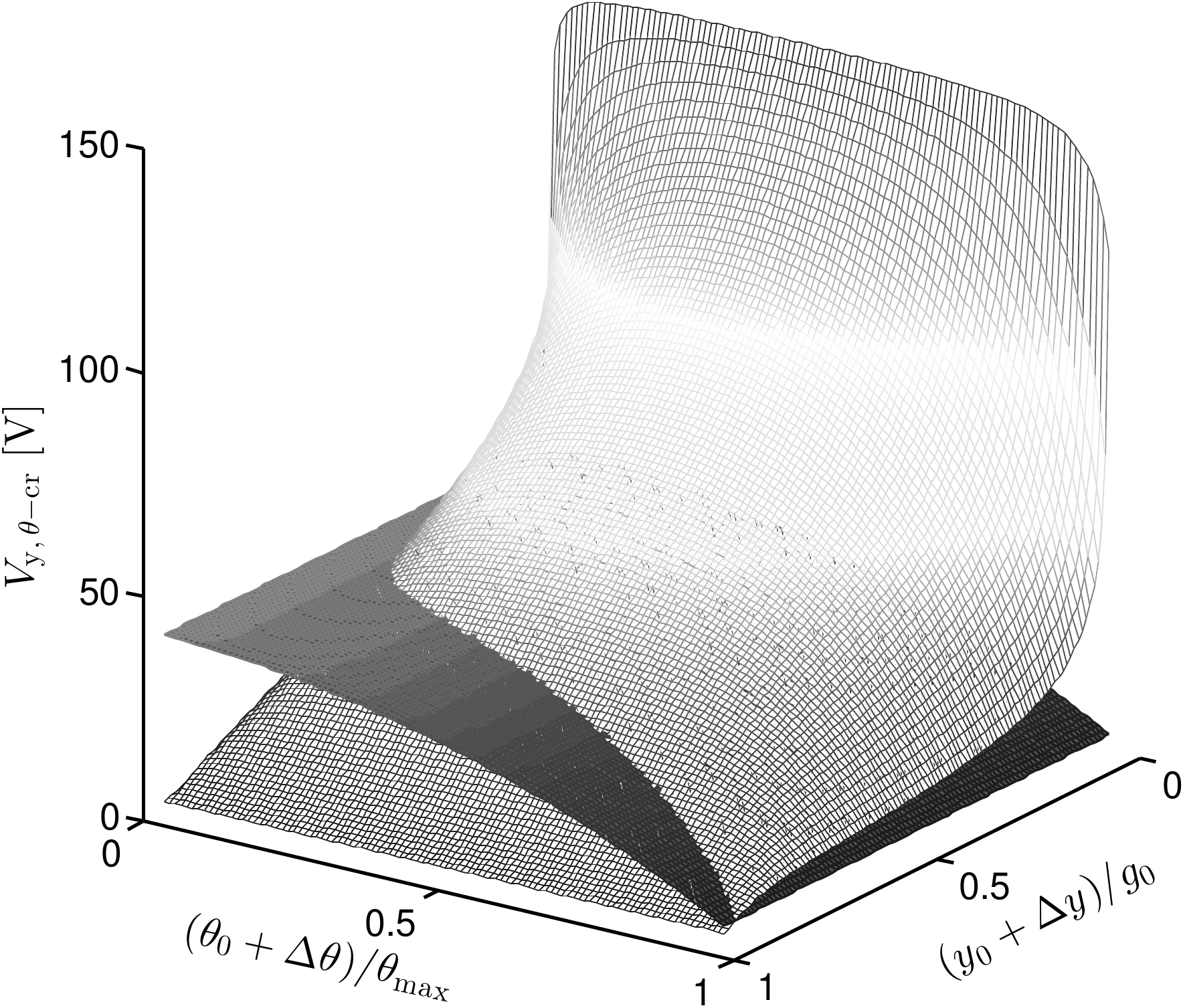}
	\caption{The intersection of two surfaces versus variation of the translational and rotational displacement, which determines the critical voltage.}
	\label{Fig:V_cr_3D}
\end{figure}
Figure \ref{Fig:V_cr_3D} presents the 3D curve of the critical voltage determined by the intersection of two surfaces in the right-hand side of the equation system \eqref{V_cr_Xmax}.

The critical voltage is now expressed as
\begin{align}
V_\mathrm{y, \theta - cr}^ \ast = \displaystyle \max \left \{ V \in \overline{V}^\ast \right\}
\end{align}
where $\overline{V}^\ast$ is a set of $V$ such that the equation \eqref{Eq:Matrix_Offsets} has solutions $\Delta y \in D_\mathrm{\Delta y}^ \ast$ and $\Delta \theta \in D_\mathrm{\Delta \theta}^ \ast$, with $D_\mathrm{\Delta y}^ \ast : \left \{ -g_0 -y_0 < \Delta y < g_0 - y_0 \right \}$ and $D_\mathrm{\Delta \theta}^ \ast : \left \{ -\theta_\mathrm{max} - \theta_0 < \Delta \theta < \theta_\mathrm{max} - \theta_0,\, \theta_\mathrm{max}=\tan^{-1} \frac{g_0}{L} \right \}$. As aforementioned, $V_\mathrm{y, \theta - cr}^ \ast$ can be solved numerically by utilizing the nonlinear constrained optimization methods such as IP or SQP. To exhibit the effect of offsets on the critical voltage, two special cases where $(\theta_0 = 0,\, y_0  \neq 0) $ or $(\theta_0  \neq 0,\,y_0 = 0)$ are separately considered.

\begin{figure}[!tbp]
	\begin{center}
		\subfigure[$V_\mathrm{y,\theta-cr}^\ast (y_0 \neq 0, \theta_0 = 0)$]{%
			\includegraphics[width=0.45\textwidth]{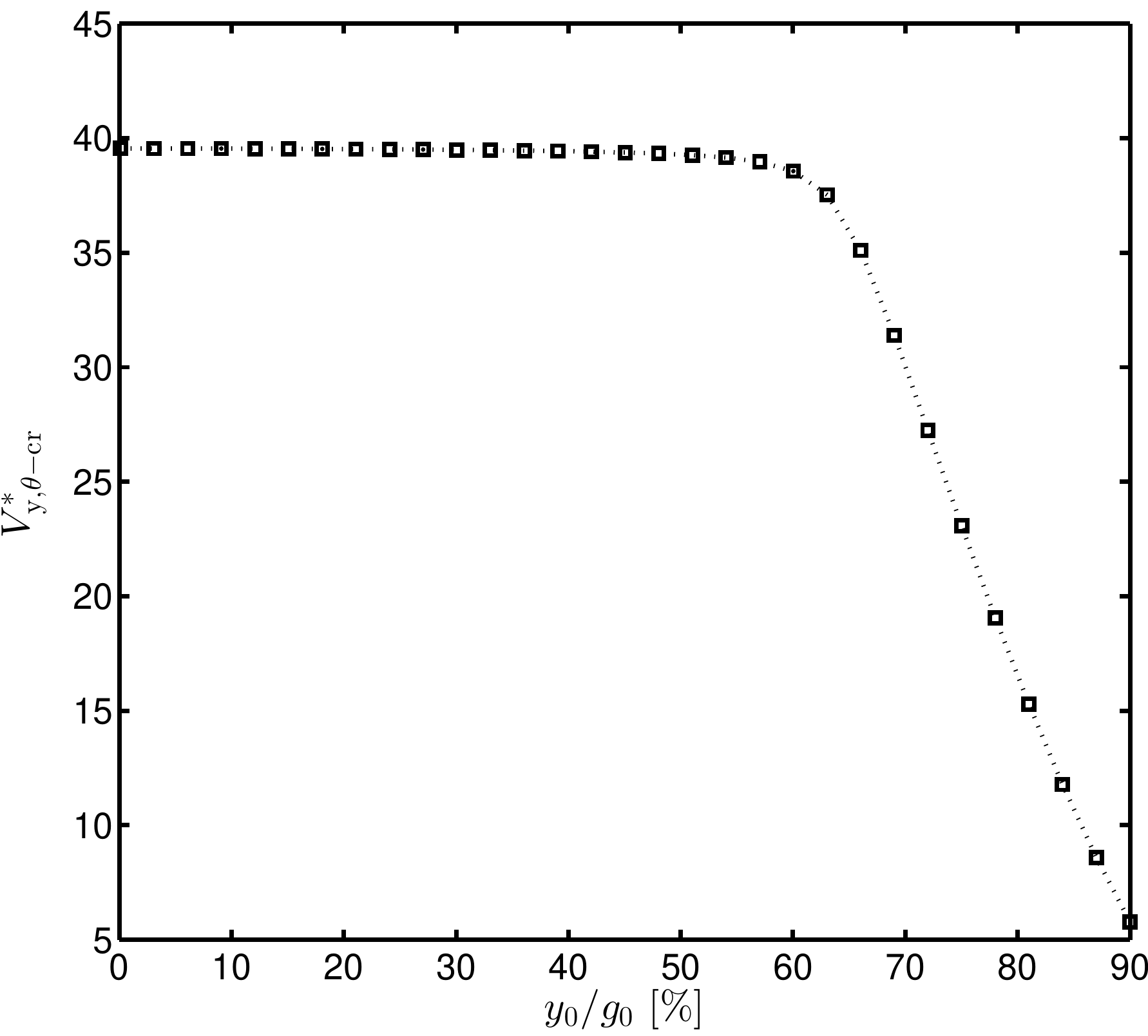}
		}%
		\space
		\subfigure[$V_\mathrm{y,\theta-cr}^\ast (y_0 = 0, \theta_0 \neq 0)$]{%
			\includegraphics[width=0.45\textwidth]{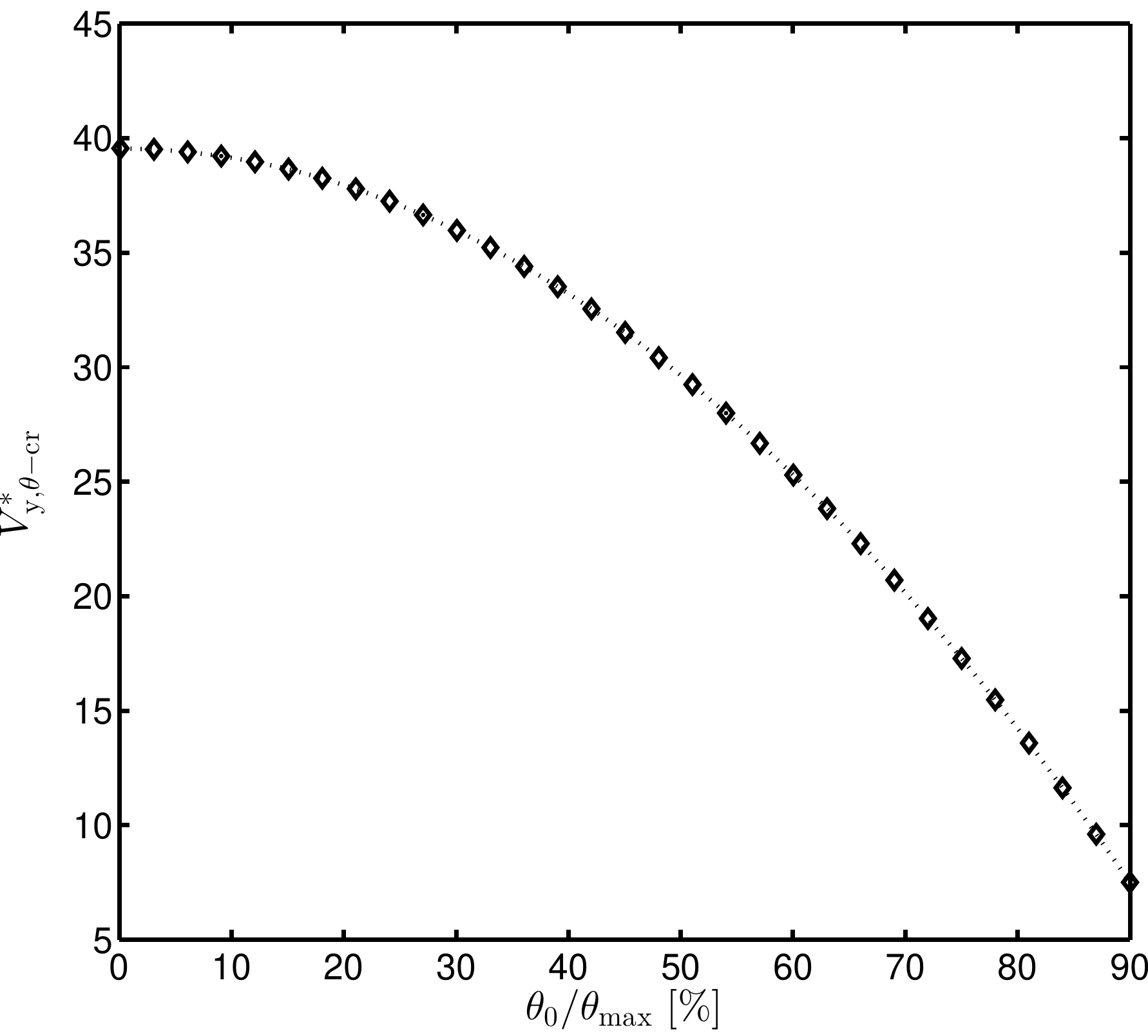}
		}
	\end{center}
	\caption{The reduction of the critical voltage taking into account the effects of misalignment offsets.}
	\label{Fig:V_cr_Offests}
\end{figure}
Figure \ref{Fig:V_cr_Offests} depicts numerical solutions of the critical voltage when the translational and rotational offsets are taken into account.
In general trend, the larger the lateral off-sets, the bigger the critical voltage reduction.
When $\theta_0 = 0$, the critical voltage $V_\mathrm{y, \theta - cr}^ \ast$ is almost unchanged if the ratio $\frac{y_0}{g_0} \leq 0.6$ and dramatically reduces with further increase of $\frac{y_0}{g_0}$. In case of $y_0 = 0$, $V_\mathrm{y, \theta - cr}^ \ast$ gradually decreases with rise of $\frac{\theta_0}{\theta_\mathrm{max}}$.

\section{Analysis of a comb-drive harvesters with two anti-phase capacitors}

\subsection{Differential common modes}
\begin{figure}[!tbp]
	\centering
	\includegraphics[width=0.28\textwidth]{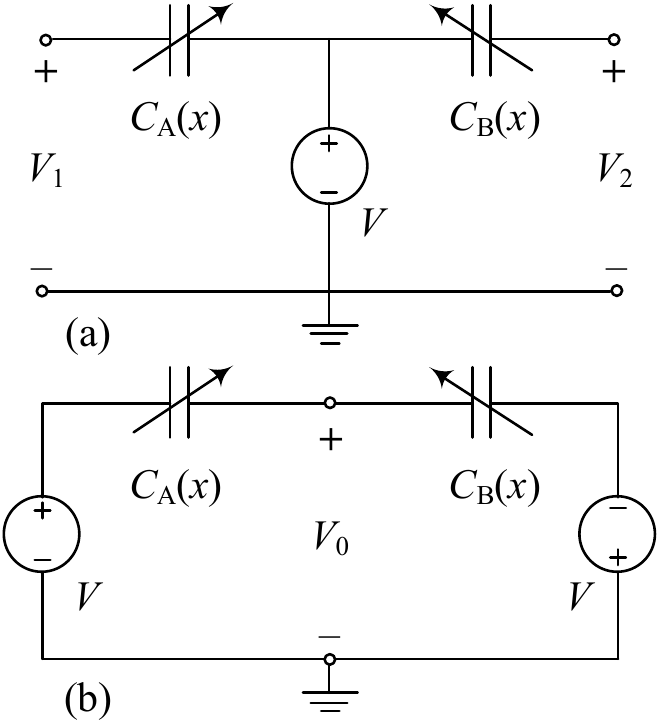}
	\caption{Circuit diagram for two common configurations of overlap-varying anti-phase transducers.}
	\label{Fig:Typical_Config}
\end{figure}
For the versatility, the overlap-varying anti-phase transducers are used in many applications, such as sensing and actuating \cite{Hirano1992, Yu2010, Chang2012, Hou2014}.
Consider common configurations of such structure represented in Figure \ref{Fig:Typical_Config}, the electrostatic energy is
\begin{align}
\small
W_\mathrm{e} = -\frac{1}{2} \big(C_\mathrm{A} + C_\mathrm{B}\big) V^2
\end{align}
where $C_\mathrm{A} = C_1 + C_2$ and $C_\mathrm{B} = C_3 + C_4$. 
$C_1$ and $C_2$ are referred to \eqref{C1_gen} and \eqref{C2_gen}, while $C_3$ and $C_4$ are calculated as
\begin{align}
\small
C_3  \big(x, y,  \theta \big) &= C_0  \frac{g_0}{2 x_0 \sin \theta} \ln \frac{g_0 + y + L \sin \theta}{g_0 + y +  \big(L -  \big(x_0 - x\big) \big)  \sin \theta} , 
\end{align}
\begin{align}
\small
C_4  \big(x, y,  \theta \big) &= C_0  \frac{g_0}{2 x_0 \sin \theta} \ln \frac{g_0 - y - \big(L -  \big(x_0 -x\big) \big) \sin \theta}{g_0 - y - L \sin \theta} .
\end{align}
The coefficients of the stiffness matrix in \eqref{Eq:Matrix} can be found in Appendix \ref{Appendix_AntiPhase}.

\subsection{Bennet's doubler configuration}

\begin{figure}[!tbp]
	\centering
	\includegraphics[width=0.35\textwidth]{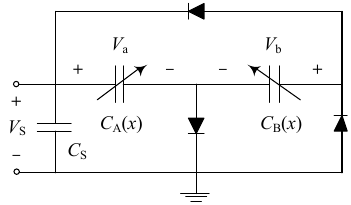}
	\caption{Bennet's doubler configuration of the overlap-varying anti-phase transducers.}
	\label{Fig:Doubler_Config}
\end{figure}

We are now widening the problem of lateral instability for more general circumstance where the voltages $V_\mathrm{a}$ and $V_\mathrm{b}$ across $C_\mathrm{A}$ and $C_\mathrm{B}$ respectively are not equal. To be specific, the overlap-varying transducers configured as Bennet's doubler represented in Figure \ref{Fig:Doubler_Config} is investigated. The analytical solution utilizing mathematically idealized diode model in \cite{Truong2017c} shows that $V_\mathrm{a}$ and $V_\mathrm{b}$ can be captured by DC offset sinusoidal signals when the doubler circuit reaches saturation. For the static pull-in instability analysis, the DC offset voltages on $C_\mathrm{A}$ and $C_\mathrm{B}$ are considered and respectively expressed as
\begin{align}
\small
V_\mathrm{A} &= V_\mathrm{s} \frac{1 + \sqrt{5}}{4} , \\
V_\mathrm{B} &= V_\mathrm{s} \frac{3 + \sqrt{5}}{4} 
\end{align}
where $V_\mathrm{s}$ is the saturation DC voltage at output. The electrostatic energy is
\begin{align}
\small
W_\mathrm{e} = -\frac{1}{2} V_\mathrm{s}^2 \big( \frac{3+\sqrt{5}}{8} C_\mathrm{A} + \frac{7+3\sqrt{5}}{8} C_\mathrm{B}\big) .
\end{align}
Similarly, the complete global stiffness matrix can be obtained by taking the derivative of the moment and forces, see Appendix \ref{Appendix_Doubler} for more details.

\begin{figure}[!tbp]
	\begin{center}
		\subfigure[Anti-phase operation mode]{%
			\includegraphics[width=0.45\textwidth]{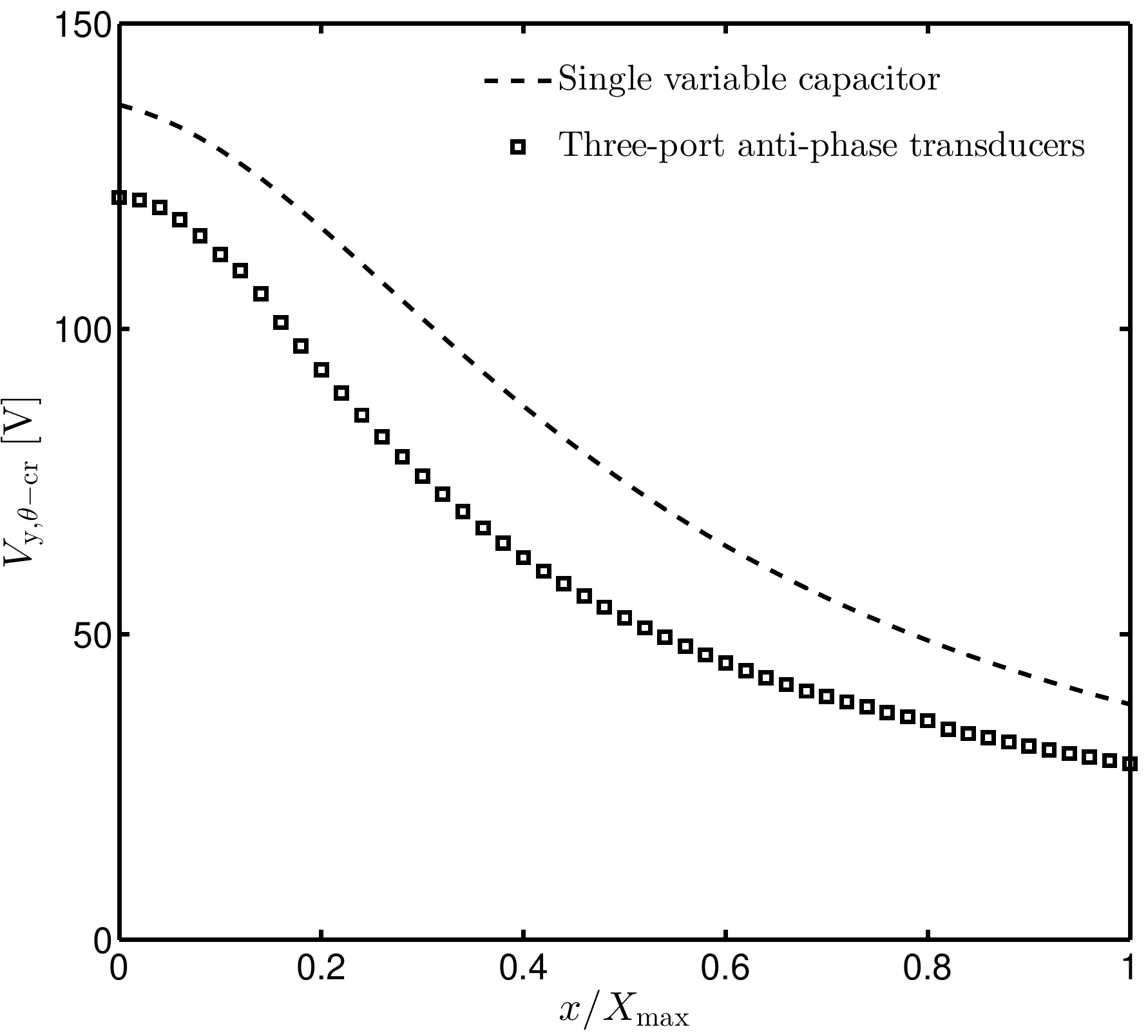}
		}%
		\space
		\subfigure[Bennet's doubler configuration]{%
			\includegraphics[width=0.45\textwidth]{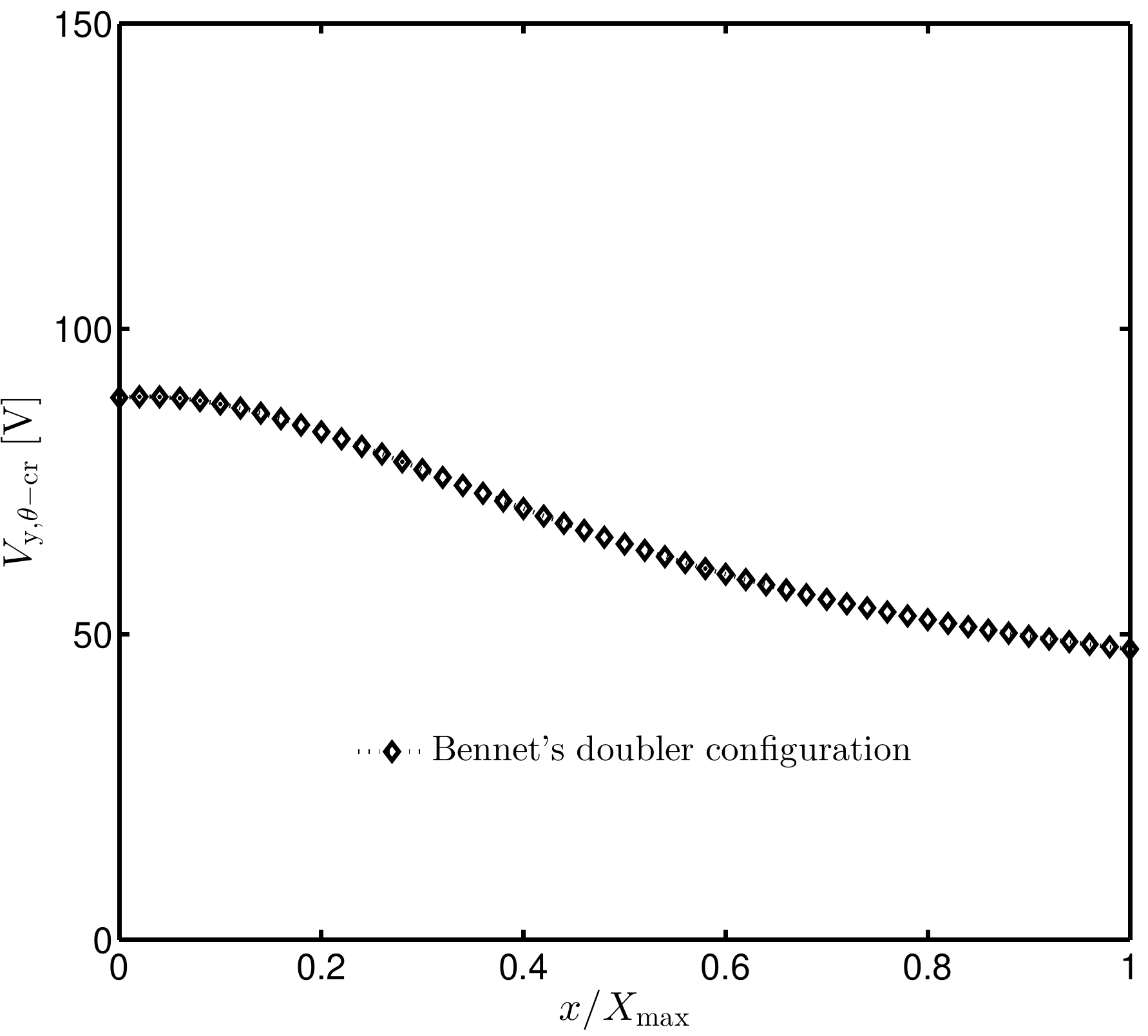}
		}
	\end{center}
	\caption{Comparison of the critical voltages for the single-capacitance transducer and different configurations of the one with anti-phase capacitors.}
	\label{Fig:V_cr_2trans}
\end{figure}

\subsection{Numerical results}
Using the same optimization procedure as presented in previous section, the critical voltage is numerically obtained in Figure \ref{Fig:V_cr_2trans}. For the common configurations shown in Figure \ref{Fig:Typical_Config}, there is a significant reduction of $V_\mathrm{y,\,\theta - cr}$, i.e. more than 10 V, comparing to the case that the single-capacitance transducer is investigated. $V_\mathrm{y,\,\theta - cr}$ of the doubler circuit also decreases with increase of the proof mass displacement, but less sensitive than the two former circumstances.

\section{Discussion}

In principle, pull-in phenomenon is the loss of the equilibrium stability, from which one should distinguish the difference between the static and dynamic pull-in aspects.
The static conditions based on potential energy are for local stability, they are only applicable small displacement near the equilibrium point.
Younis \cite{Younis2011} presented an universal definition of dynamic pull-in, which is the collapse of the movable electrode into the stationary one due to the combined action of the kinetic and potential energies. For the considered transducers, source of kinetic energy is from the AC harmonic voltages. The dynamic pull-in generally occurs at lower DC voltage compared to that of static pull-in, see \cite{Elata2006} for an example. Dynamic pull-in instability therefore can be considered as a key source of failure
in MEMS electrostatic devices. It is more of interest for sophisticated configurations that have been studied, and is an open issue for further investigations.

When the overlap-varying transducers are configured as Bennet's doubler, the max/min ratio of capacitance variation needs to be larger than 2 to allow operation of the circuit \cite{deQueiroz2011}. The travel range of the proof mass is now more important, which is fundamentally dictated by inherent pull-in instability. In attempts to enlarge the stable displacement for MEMS electrostatic devices, several improvements of the suspension beam designs have been developed. Zhou \textit{et al.} proposed a tilted folded-beam suspension to shift the maximum of the lateral spring constant curve and thus prevent the pull-in limited travel range of the comb-drives \cite{Zhou2003}. For more recent work, Olfatnia \textit{et al.} presented a novel clamped paired double parallelogram flexure mechanism, offering high stroke direction stiffness $K_x$ while maintaining low translational and rotational stiffnesses $K_y$ and $K_\theta$ over a large range of proof mass displacement \cite{Olfatnia2013}. These advanced methods can be extremely useful to overcome the challenging in enhancing the stable range.

\section{Conclusion}

This paper presents a further development of a 2D model utilizing to analyze the lateral pull-in instability of an in-plane overlap-varying transducer. Analytical solutions of the critical voltage are derived when the translational and rotational displacements are separately considered. 
The ratio of rotational and translational critical voltages in these two cases can be made large by appropriate choice of the dimensions $D_1$ and $D_2$ of the folded-beam spring.
The numerical result for the general case is determined taking into account combination of both lateral translation and rotation. The effects of translational and rotational offsets to the critical voltage are explored numerically.
All analysis results are adaptable and applicable to different type of the mechanical spring, and therefore can be used as a guideline for MEMS transducer design.

%
%
\vfill

\small
\section*{Acknowledgement}
\small
This work was supported by the Research Council of Norway through Grant no. 229716/E20.


\appendix
\section{Anti-phase operation mode} \label{Appendix_AntiPhase}
With $\sin \theta \approx \theta$, the coefficients of the stiffness matrix in \eqref{Eq:Matrix} would take the forms

\begin{align}
\small
\begin{split}
\frac{\partial F_\mathrm{y}}{\partial y} \at[\Big]{y \rightarrow 0} = -K_y + \frac{1}{4} V^2 \frac{C_0 g_0}{x_0} \left[ \frac{x_0 + x}{(g_0 + L\theta) \big( g_0 + \big( L - (x_0 +x) \big) \theta \big)^2} + \frac{x_0 + x}{(g_0 + L\theta)^2 \big( g_0 + \big( L - (x_0 +x) \big) \theta \big)} \right. \\
+ \frac{x_0 + x}{(g_0 - L\theta) \big( g_0 - \big( L - (x_0 +x) \big) \theta \big)^2} + \frac{x_0 + x}{(g_0 - L\theta)^2 \big( g_0 - \big( L - (x_0 +x) \big) \theta \big)} \\
+ \frac{x_0 - x}{(g_0 + L\theta) \big( g_0 + \big( L - (x_0 - x) \big) \theta \big)^2} + \frac{x_0 - x}{(g_0 + L\theta)^2 \big( g_0 + \big( L - (x_0 - x) \big) \theta \big)} \\
\left. + \frac{x_0 - x}{(g_0 - L\theta) \big( g_0 - \big( L - (x_0 - x) \big) \theta \big)^2} + \frac{x_0 - x}{(g_0 - L\theta)^2 \big( g_0 - \big( L - (x_0 - x) \big) \theta \big)} \right] ,
\end{split}
\end{align}
\begin{align}
\small
\frac{\partial F_\mathrm{y}}{\partial \theta} \at[\Big]{\theta  \rightarrow 0} = - V^2 \frac{C_0 g_0^2}{x_0} \frac{\big(g_0^2 + 3y^2\big) \big(x_0^2 + x^2 - 2L x_0 \big)}{\big(g_0 - y\big)^3 \big(g_0 + y\big)^3} ,
\end{align}
\begin{align}
\small
\begin{split}
\frac{\partial M_\mathrm{\theta}}{\partial y} \at[\Big]{y  \rightarrow 0} = \frac{1}{4} V^2 \frac{C_0 g_0 }{\theta x_0} \left[ -\frac{x_0+x}{( g_0-L \theta) \big(g_0-\big (L -(x_0+x) \big) \theta \big)} - \frac{x_0-x}{(g_0-L \theta) \big(g_0- \big(L -(x_0 - x) \big)\theta\big)} \right. \\
+\frac{x_0-x}{(g_0+L \theta) \big (g_0+\big(L - (x_0 -x) \big) \theta \big)}+\frac{x_0+x}{(g_0+L \theta) \big(g_0+ \big(L -(x_0 + x)\big) \theta \big)} \\
+\frac{(x_0+x) g_0}{(g_0-L \theta)^2 \big(g_0-\big(L -(x_0-x)\big) \theta \big)} + \frac{(x_0-x) g_0}{(g_0-L \theta)^2 \big(g_0- \big(L -(x_0-x)\big) \theta \big)}  \\
+ \frac{L \theta (x_0-x)}{(g_0+L \theta)^2 \big(g_0+\big(L - (x_0 - x)\big) \theta \big)}+\frac{L \theta (x_0+x)}{(g_0+L \theta)^2 \big(g_0+\big(L-(x_0+x) \big) \theta \big)} \\
+\frac{\theta (x_0+x) \big(L-(x+x_0)\big)}{(g_0-L \theta) \big(g_0-\big(L -(x_0+x) \big) \theta \big)^2}+\frac{\theta (x_0-x) \big(L-(x_0-x)\big)}{(g_0-L \theta) \big(g_0-\big(L -(x_0-x) \big) \theta\big)^2} \\
\left. -\frac{g_0 (x_0-x)}{(g_0+L \theta) \big(g_0+ \big(L -(x_0-x) \big) \theta \big)^2} -\frac{g_0 (x_0+x) }{(g_0+L \theta) \big(g_0+ \big(L -(x_0+x) \big) \theta\big)^2}\right] ,
\end{split}
\end{align}
\begin{align}
\small
\frac{\partial M_\mathrm{\theta}}{\partial \theta} \at[\Big]{\theta  \rightarrow 0} = - K_\theta + \frac{2}{3} V^2 \frac{C_0 g_0^2}{x_0} \frac{\big(g_0^2 + 3y^2\big) \big( 3L^2 x_0 - 3L^2 x^2- 3L x_0^2 + 3 x^2 x_0 +x_0^3 \big)}{\big(g_0 - y\big)^3 \big(g_0 + y\big)^3} .
\end{align}

\vfill

\section{Bennet's doubler configuration} \label{Appendix_Doubler}
For the doubler configuration, the coefficients of the stiffness matrix in \eqref{Eq:Matrix} are

\begin{align}
\small
\begin{split}
\frac{\partial F_\mathrm{y}}{\partial y} \at[\Big]{y \rightarrow 0} = -K_y + V^2 \frac{C_0 g_0^2 }{x_0 \left(g_0^2-L^2 \theta^2\right)^2} \\
\left[ \frac{p (x_0-x) \left(g_0^4+2 g_0^2 L \theta^2 (L-(x_0-x))-L \theta^4 \left(3 L^3-6 L^2 (x_0-x)+4 L (x_0-x)^2-(x_0-x)^3\right)\right)}{\left(g_0^2-\theta^2 \left(L-(x_0-x) \right)^2\right)^2} \right.\\
\left.  +\frac{q (x_0+x) \left(g_0^4+2 g_0^2 L \theta^2 (L-(x_0+x))+L \theta^4 \left(-3 L^3+6 L^2 (x_0+x)-4 L (x_0+x)^2+(x_0+x)^3\right)\right)}{\left(g_0^2-\theta^2 \left(L-(x+x_0)\right)^2\right)^2} \right] ,
\end{split}
\end{align}

\begin{align}
\small
\begin{split}
\frac{\partial F_\mathrm{y}}{\partial \theta} \at[\Big]{\theta  \rightarrow 0} = -\frac{1}{2} V^2 \frac{C_0 g_0^2 \left(g_0^2+3 y^2\right) \left(-2 L (p (x_0-x)+q (x_0+x))+p (x_0-x)^2+q (x_0+x)^2\right)}{x_0 \left(g_0^2-y^2\right)^3} ,
\end{split}
\end{align}

\begin{align}
\small
\begin{split}
\frac{\partial M_\mathrm{\theta}}{\partial y} \at[\Big]{y  \rightarrow 0} = \frac{1}{2} V^2 \frac{C_0 g_0^2 }{x_0} \left[\frac{2 p (x_0-x) \big(2 L-(x_0-x)\big) \left(g_0^4-L^2 \theta^4 \big(L-(x_0-x)\big)^2\right)}{\left(g_0^2-L^2 \theta^2\right)^2 \left(g_0^2-\theta^2 \big(L-(x_0-x)\big)^2\right)^2} \right. \\
+\frac{2 q (x_0+x) \big(2 L-(x_0+x) \big) \left(g_0^4-L^2 \theta^4 \big(L-(x_0+x) \big)^2\right)}{\left(g_0^2-L^2 \theta^2\right)^2 \left(g_0^2-\theta^2 \big(L-(x_0+x)\big)^2\right)^2} \\
\left. -\frac{p (x_0-x) \big(2 L-(x_0-x)\big)}{\left(g_0^2-L^2 \theta^2\right) \left(g_0^2-\theta^2 \big(L-(x_0-x)\big)^2\right)} 
- \frac{q (x_0+x) \big(2 L-(x_0+x) \big)}{\left(g_0^2-L^2 \theta^2\right) \left(g_0^2-\theta^2 \big(L-(x_0+x)\big)^2\right)}\right] ,
\end{split}
\end{align}

\begin{align}
\small
\begin{split}
\frac{\partial M_\mathrm{\theta}}{\partial \theta} \at[\Big]{\theta  \rightarrow 0} = - K_\theta + \frac{1}{3} V^2 \frac{C_0 g_0^2 \left(g_0^2+3 y^2\right) }{x_0 \left(g_0^2-y^2\right)^3} \left[3 L^2 (p (x_0-x)+q (x_0+x)) \right. \\
\left. -3 L \left(p (x_0-x)^2+q (x_0+x)^2\right)+p (x_0-x)^3+q (x_0+x)^3\right] 
\end{split}
\end{align}
where 
\begin{align}
\small
p &= \frac{1 + \sqrt{5}}{4} , \\
q &= \frac{3 + \sqrt{5}}{4} .
\end{align}

\vfill


\end{document}